\documentstyle[prl,preprint,tighten,aps]{revtex}
\begin{document}

\title{Average over energy effect of parity nonconservation in neutron
scattering on heavy nuclei }

\author {O. P. Sushkov$^{*}$ }
\address { School of Physics, The University of New South Wales,
                Sydney 2052, Australia}

\maketitle

\begin{abstract}
Using semiclassical approximation we consider parity nonconservation 
(PNC) averaged over compound resonances. We demonstrate that the result
of the averaging crucially depends on the properties of residual
strong nucleon-nucleon interaction. Natural way to elucidate this
problem is to investigate experimentally PNC spin rotation with
nonmonachromatic neutron beam: $E \sim \Delta E \sim 1MeV$. Value
of the effect can reach $\psi \sim 10^{-5}-10^{-4}$ per mean free
path.
\end{abstract}


Enhancement of parity nonconservation (PNC) in neutron
scattering near compound resonances was predicted in  1980\cite{Sus}
and first observed experimentally in  Dubna in 1981\cite{Alf} for
lowest p-wave resonance in $^{117}$Sn. During next decade the effect 
was observed in several more resonances in different nuclei. The
magnitudes of the effect were in brilliant agreement with statistical
theory\cite{Sus} (for review of theory see articles\cite{Sus1,Flam}).
New development arose in 1990-1991 with appearance of Los Alamos data
for $^{238}$U\cite{Bow} and especially for $^{232}$Th\cite{Fran}
(also see paper\cite{Fran1}). Number of resonances where the
effect was observed became big, and it was clear that average
value of the effect was not equal to zero, contrary to the expectation 
of the statistical theory. There have been several theoretical
works devoted to this problem (for a review see Ref.\cite{Flam}),
but situation remains unclear.

 Let us first consider possible scenarios for for the resolution of
the problem. 1) Nonzero average is just a statistical fluctuation,
because number of resonances where the effect has been observed
is still not very big. 2) There is some gross structure in the
effect: nonzero average over interval $\epsilon$ which is much
bigger than distance between resonances and much smaller than
1 MeV ($D \ll \epsilon \ll \Delta E \sim 1 MeV$). But the average 
over interval $\Delta E \sim 1 MeV$ is still zero or at least
very small. 3) Nonzero average over $\Delta E \sim 1 MeV$. 

First possibility can not be ruled out, but  even if present
value is a fluctuation there is a question about expected
average value.
Existence of gross structure would mean that there are some
intermediate states of opposite parity separated by the interval
$\epsilon$. This relatively small interval can work only if it is
bigger than the  width of corresponding states. However the spread
widths of any known intermediate state $\Gamma_{spr} \sim 1 MeV$,
and this probably kills the gross structure scenario. In the present work
we consider third scenario: Nonzero average over interval
$\Delta E \sim \Gamma_{spr} \sim 1MeV$.  It will be important for
us that only elastic channel is open [$(n,\gamma)$ can be 
neglected anyway] . This is why $\Delta E$ should not be bigger
than 1 MeV.

Thus let us consider scattering of a neutron with energy
$E \sim \Delta E \sim 1 MeV$. Certainly at this energy there
is no kinematic enhancement of PNC effect because $kR \sim 1$
($k=\sqrt{2 m E}$ and  $R$ is radius of the nucleus). But this
is irrelevant to the problem of averaging because the kinematic
enhancement at small energy is due just to trivial suppression
of an incident neutron wave function by centrifugal barrier.

At $E \sim 1MeV$ the width of each particular compound resonance
$\Gamma_c$ is of the order of distance between the resonances
$\Gamma_r \sim D$. Therefore an averaging over the resonances is
equivalent to the averaging over continuous energy. Due to Heisenberg
uncertainty relation $\Delta E \cdot \Delta t \sim 1$ we can
replace the averaging over energy interval $\Delta E$ by
consideration of  a wave packet localized in time with
uncertainty $\Delta t \sim 1/MeV$. This is equivalent to
semiclassical approximation. Certainly at $E \sim \Delta E
\sim 1 MeV$ semiclassical parameter is not very good:
$2R/v\Delta t \sim kR \sim 1$, but still it is reasonable.
So instead of energy representation we intend to consider
space-time picture of the scattering. 

An effective Hamiltonian of a nucleon weak interaction with 
nucleus is of the form 
\begin{equation}
\label{HW}
H_W=g_a{{G}\over{2\sqrt{2}m}}\{{\bf \sigma}{\bf p},\rho(r)\}
\approx g_a{{G\rho}\over{\sqrt{2}m}}{\bf \sigma}{\bf p}.
\end{equation}
Here $G\approx 10^{-5}/m^2$ is Fermi constant, $m, {\bf p},
{\bf \sigma}$ - mass, momentum and spin of the nucleon.
$\rho$ is nuclear density ($\int \rho dV =A$) which is taken to be
constant. $g_a$ is  effective constant of the weak interaction:
$|g_n| < 1$, $g_p \approx 5$ (see e.g. Refs.\cite{Khr,Sus2}).

Let us  first consider ``direct'' or ``potential'' process
when neutron passes nucleus without excitation of other nucleons.
Angle of the neutron spin rotation is equal to the phase
difference between positive and negative helicity neutrons
\begin{equation}
\label{psi}
\psi=p_+d-p_-d=\left(\sqrt{2m(U+W_n)}-\sqrt{2m(U-W_n)}\right)d
\end{equation}
Here $U \approx 40 MeV$ is value of the nuclear potential,
and $W_n=g_nG\rho p/\sqrt{2}m=g_nG\rho \sqrt{U/m}$
is  weak interaction. Assuming that $d \sim R=r_0 A^{1/3}$
we find from (\ref{psi})
\begin{equation}
\label{psi1}
\psi \sim \sqrt{2} g_n R G \rho =g_n{3\over{2\sqrt{2}\pi}}
10^{-5}{{A^{1/3}}\over{m^2r_0^2}} \approx g_n 10^{-6}.
\end{equation}
Per mean free pass each neutron passes in average one nucleus.
It means that above estimation corresponds to the neutron mean 
free path.

During passage of nucleus the neutron can excite other nucleons
and hence produce the cascade. Probability of this process is 
about 1, because at spread width $\Gamma_{spr} \sim 3MeV$ mean
free path of a neutron in nuclear matter is of the order of
nuclear size: $l/2R \sim \sqrt{2U/m}/2\Gamma_{spr} R \approx 1.5$
The nucleons excited in the cascade can not get out of nucleus
because energy of each particular  nucleons is not enough to
escape. So nucleons are trapped. If nucleus would be an infinite
system the cascade would never inverted due to second law of
thermodynamics. However nucleus is finite with finite density of
spectrum, therefore the cascade is inverted after quantum time
$\tau \sim 1/D$, and neutron escapes.
Ratio of the neutron life time in this quantum trap to the
time of free pass is 
\begin{equation}
\label{N}
N={v\over{2DR}}={{\sqrt{2U/m}}\over{2D r_0 A^{1/3}}}
\sim {{5MeV}\over{D}} \sim 10^6.
\end{equation}
This is exactly the parameter which involves in the compound
resonance theory\cite{Sus}.

Life time in the trap can enter into some physical effects.
For example let us imaging some magnetic field inside nucleus
and assume that nuclear forces are spin independent. Then
Faraday rotation of the spin of the trapped neutron is
$\mu_n{\cal H} \tau$, and this is by $N$ times bigger than
the Faraday rotation in the ``potential'' process.
(One can easily come to the same conclusion using perturbation
theory and compound-resonance representation).

We are interested in the weak interaction (\ref{HW}) which is 
proportional to ${\bf \sigma}{\bf p}$. Let us assume first that
nuclear forces (residual interaction, as well as selfconsistent 
potential) are spin independent. In this situation spin is 
separated from orbital motion, and average value of the momentum
$p$ in cascade corresponds to free passage. However, due to the
charge exchange forces, the cascade protons also contribute
into spin rotation, and estimation (\ref{psi1}) should be
replaced by
\begin{equation}
\label{psi2}
\psi \sim g_p 10^{-6} \sim 0.5 10^{-5}.
\end{equation}
Let us assume now that helicity of the particle is conserved
in the nuclear scattering. The examples of such conservation
are well known. It is enough to recall that in quantum 
electrodynamics at high energy scattering the helicity of an
electron is conserved. With the assumption about helicity
conservation we come to the estimation of the neutron spin
rotation $\psi \sim N g_p 10^{-6} \sim g_p$.  It  certainly
is not realistic. Helicity in the cascade is not conserved
exactly. Let us denote by $\tau_m$ the time of memory about
initial helicity. Then we get the following estimation of
the angle of neutron spin rotation.
\begin{equation}
\label{psi3}
\psi \sim {{\tau_m}\over{\tau_f}}g_p 10^{-6}.
\end{equation}
Here $\tau_f \sim 2R/v$ is the time of free passage of nucleus.
To fit experimental data we need $\tau_m/\tau_f \sim 10$
(One should not forget about kinematic enhancement 
$\sim 1/kR \sim 5 \cdot 10^2$ for low energy data\cite{Bow,Fran,Fran1}).
We would like to note that the considered picture is somewhat similar
to the ``quasielastic mechanism'' suggested in Ref.\cite{Flam1}
(see also Ref.\cite{Hus}).
However it is probably impossible to derive semiclassical
result basing on perturbation theory used in Refs.\cite{Flam1,Hus}.

After first step of the cascade we have three quasiparticles
in the system.  So each quasiparticle has in average 1/3 of the
initial excitation energy. According to standard Landau estimation,
the spread width of quiasiparticle is proportional to $\propto E^2$.
Therefore $\Gamma_{spr}^{\prime} \sim (1/3)^2\Gamma_{spr} \sim 0.3 MeV$.
It means that each of these quasiparticles lives long enough
before decay into more complicated configurations, and in first
approximation we can forget about this decay (cf. with Ref.\cite{Hus}),
but basically it should be included as well.  To avoid misunderstanding
we stress that $\Gamma_{spr}^{\prime}$ is the spread width of qusiparticle.
The spread width of configuration is 3 times bigger. There is a bunch
of classical trajectories in nuclear potential along which
the helicity is conserved. However to simulate the cascade
one needs also to know the spin-isospin structure of the
residual strong interaction. Unfortunately, existing 
parametrizations\cite{Mig} are not very reliable. Nevertheless
one can try to do Monte Carlo simulations of the cascade.
We can move in  opposite direction and consider the experimental
data\cite{Bow,Fran,Fran1} as a direct measurement of the memory time:
$\tau_m/\tau_f \sim 10$. In this case we can expect the angle of
neutron spin rotation per mean free pass of nonmonochromatic
beam ($\Delta E \sim 1MeV$) to be $\psi \sim 10^{-4}$.
The discussed effect is independent of nucleus and in this sense
it is regular, but still it is related to the complex structure 
of the cascade and sign of the effect is not obvious.

In the present paper we have considered semiclassical picture
of the neutron spin rotation caused by parity nonconserving
weak interaction. We demonstrated that the calculation can be
reduced to Monte Carlo simulation of the  cascade
which is much simpler than an exact quantum calculation for
compound states. The semiclassical picture is equivalent to
the averaging of the effect over energy interval $\Delta E \sim 1MeV$.
This can be investigated experimentally with nonmonochromatic
neutron beam. We  demonstrated that the angle of rotation can reach
$\psi \sim 10^{-5} \ - \ 10^{-4}$ per mean free path. The
upper estimation ($\psi \sim 10^{-4}$) is based on experimental
data\cite{Bow,Fran,Fran1} when we treat them  as a measurement of 
$\tau_m$.

\vskip3ex 
I am very grateful  to V. V. Flambaum, G. F. Gribakin and 
D. L. Shepelyansky for discussions.
This  work has been completed during my stay at the Laboratoire de 
Physique Quantique, Universite Paul Sabatier. I am gratefully 
acknowledge the hospitality and financial support.



\begin{references}
\bibitem[*]{budker} Also at the Budker Institute of Nuclear Physics,
              630090 Novosibirsk, Russia.
\bibitem{Sus} O. P. Sushkov and V. V. Flambaum, Pis'ma Zh. Eksp. Teor.
Fiz., {\bf 32}, 377 (1980) [JETP Lett., {\bf 32}, 363 (1980)].
\bibitem{Alf} V. P. Alfimenkov et al, Pis'ma Zh. Eksp. Teor.
Fiz., {\bf 34}, 308 (1981) [JETP Lett., {\bf 34}, 295 (1981)].  
\bibitem{Sus1} O. P. Sushkov and V. V. Flambaum,  Usp. Fiz. Nauk,
{\bf 136}, 3 (1982) [Sov. Phys.  Usp., {\bf 25}, 1 (1982)].
\bibitem{Flam} V. V. Flambaum and G. F. Gribakin, Prog. Part. Nucl.
Phys., {\bf 35}, 423 (1995).
\bibitem{Bow} J. D. Bowman et al, Phys. Rev. Lett., {\bf 65},
1192 (1990).
\bibitem{Fran} C. M. Frankle et al, Phys. Rev. Lett., {\bf 67},
564 (1991).
\bibitem{Fran1} C. M. Frankle et al, Phys. Rev., C {\bf 46},
778 (1992). 
\bibitem{Khr} V. V. Flambaum, I. B. Khriplovich and O. P . Sushkov,
Phys. Lett., B {\bf 146}, 367 (1984).
\bibitem{Sus2} O. P. Sushkov and V. B. Telitsin, Phys. Rev. C
{\bf 48}, 1069 (1993).
\bibitem{Flam1} V. V. Flambaum, Phys. Rev., C {\bf 45}, 437 (1992).
\bibitem{Hus} M. S. Hussein, A. K. Kerman and C-Y Lin, Z. Phys. A
{\bf 351}, 301 (1995).
\bibitem{Mig} A. B. Migdal, Theory of Finite Fermi Systems and
Applications to Atomic Nuclei. John Wiley \& Sons, New York, 1967.
G. E. Brown, Rev. Mod. Phys., {\bf 43} 1 (1971).
V. A. Khodel and E. E. Saperstein, Phys. Rep., {\bf 92}, 183 (1982).
\end{references}
\end{document}